\documentclass[journal]{IEEEtran}
\usepackage{multirow} 
\usepackage{amsmath,amssymb} 
\ifCLASSINFOpdf
\usepackage[pdftex]{graphicx}
\usepackage{float}
\else
\fi
\usepackage{amsmath}   
\begin{document}
%
\title{Learned Lossless Image Compression Through Interpolation With Low Complexity}
%
%
%


\author{Fatih Kamisli
\thanks{The author is with the Department of Electrical and Electronics Engineering at the Middle East Technical University, Ankara, Turkey. (email: kamisli@metu.edu.tr)}
\thanks{Codes are available at https://github.com/metu-kamisli/LLICTI}}

\maketitle

\begin{abstract}
With the increasing popularity of deep learning in image processing, many learned lossless image compression methods have been proposed recently. One group of algorithms that have shown good performance are based on learned pixel-based auto-regressive models, however, their sequential nature prevents easily parallelized computations and leads to long decoding times. Another popular group of algorithms are based on scale-based auto-regressive models and can provide competitive compression performance while also enabling simple parallelization and much shorter decoding times. However, their major drawback are the used large neural networks and high computational complexity. This paper presents an interpolation based learned lossless image compression method which falls in the scale-based auto-regressive models group. The method achieves better than or on par compression performance with the recent scale-based auto-regressive models, yet requires more than 10x less neural network parameters and encoding/decoding computation complexity. These achievements are due to the contributions/findings in the overall system and neural network architecture design, such as sharing interpolator neural networks across different scales, using separate neural networks for different parameters of the probability distribution model and performing the processing in the YCoCg-R color space instead of the RGB color space.
\end{abstract}


%
\IEEEpeerreviewmaketitle

\section{Introduction} \label{sec:intro}
Most image and video communication applications use lossy compression, however, there are also applications that require lossless compression. Lossless compression allows the original data to be perfectly reconstructed from the compressed bitstream without any loss of information. For example, medical imaging, satellite imaging, professional photography and digital cinema are applications where lossless image or video compression is used. 
This paper presents a novel learned lossless image compression method.

Classical lossless image compression methods are based on pixel \cite{pennebaker1992jpeg, boutell1997png, weinberger2000loco, sneyers2016flif} or block based prediction \cite{webp, lee2006improved, zhou2012hevc}, inter-to-integer transforms \cite{calderbank1997lossless} or both \cite{kamisli2017lossless}. In the prediction based methods, the compression is typically performed in a raster scan order of pixels or blocks. A pixel or a block of pixels is predicted using previously coded left and upper neighbor pixels or neighboring left and upper pixels of the block, and the prediction error pixel or block is then entropy coded. Then the compression of the next pixel or block starts. The prediction error pixels are typically more decorrelated than the original pixels, allowing good compression performance with simple entropy codes. For example, JPEG-LS \cite{pennebaker1992jpeg} is based on a pixel based prediction algorithm and lossless compression in HEVC \cite{zhou2012hevc, alvar2016lossless} is based on a block based prediction algorithm. 

In the integer-to-integer transform based lossless compression methods, the integer-valued image pixels are transformed into a transform domain with integer-valued coefficients \cite{le1988sub, calderbank1997lossless, kamisli2017lossless}. The transformation is invertible and thus no loss of information occurs. For example, lossless compression in  JPEG2000 is based on the integer 5/3 wavelet transform \cite{le1988sub, rabbani2002overview, christopoulos2000jpeg2000}. The obtained transform coefficients are again more decorrelated than the image pixels and allow for good compression performance with simple entropy codes. It can also be useful to use the prediction and integer-to-integer transform methods together in some contexts \cite{kamisli2017lossless}.

With the high expressive power of artificial neural networks and the recent advances and interest in deep learning, lossy \cite{balle2016end, balle2018variational, minnen2018joint, minnen2020channel} and lossless \cite{mentzer2019practical, cao2020lossless, zhang2020lossless} image compression methods with neural networks have been proposed. The state of the art in learned lossy \cite{minnen2020channel, koyuncu2022contextformer} and lossless \cite{cao2020lossless, zhang2020lossless} image compression is on par with or exceeds the compression performance of state of the art classical compression methods \cite{bross2021overview, sneyers2016flif}.

Recent learned lossless image compression methods can be roughly categorized to three groups: methods based on pixel-based auto-regressive models, methods based on integer discrete flows and methods based on scale-based auto-regressive models. Methods based on pixel-based auto-regressive models are similar to the methods based on classical pixel based prediction methods in the sense that they encode/decode the image in a pixel-by-pixel manner yet use neural networks to obtain the probabilities of the pixel to be coded/decoded from previously encoded/decoded neighbor pixels. While achieving quite good compression performance, their major drawback is the long decoding times due to the sequential runs of the neural network for every pixel in the image \cite{schiopu2020deep, gumus2022learned}. Methods based on integer discrete flows are similar to classical systems based on inter-to-integer transforms as discrete flows are neural network based systems mapping integer pixels to integer latent/transform variables. Their major drawback is that they require quite big neural networks for good compression performance \cite{hoogeboom2019integer}. Methods based on scale-based auto-regressive models decompose the original image into a multiple scale representation, encode/decode first the lowest resolution scale and then encode/decode the remaining scales sequentially conditioned on previously encoded/decoded scales. These systems typically fare better than the previous two in terms of encoding/decoding speed and/or neural networks size \cite{mentzer2019practical, cao2020lossless, zhang2020lossless}.

This paper presents a learned lossless image compression algorithm that falls in the scale-based auto-regressive model category. The system first encodes/decodes a sub-sampled grid of the original image grid (e.g. every $32^{nd}$ pixel horizontally and vertically) using simple fixed-length codes. Then it progressively interpolates (i.e. produces probability distributions of) finer sub-sampled grids of the original image grid conditioned on all previously encoded/decoded grids so far, to achieve the encoding/decoding of the entire image. The interpolators are all neural network based and learned from a dataset.

The contributions of this paper are as follows. The system presented achieves better than or on par compression performance with the recent literature, yet requires more than 10x less neural network parameters and encoding/decoding computational complexity  \cite{mentzer2019practical, cao2020lossless, zhang2020lossless}.
These achievements are due to the contributions/findings in the overall system and neural network architecture design, such as 
\begin{itemize}
    \item using the same interpolator neural networks in different scales, 
    \item using separate neural networks for different parameters of the probability distribution and 
    \item performing the processing in the YCoCg-R color space instead of the RGB color space,
\end{itemize} which all help to improve compression performance and/or reduce complexity, as detailed in Section \ref{sec:prop} and \ref{sec:expres}.

The remainder of the paper is organized as follows. Section \ref{sec:relwrk} reviews recent work on learned lossless image compression with scale-based auto-regressive models. Section \ref{sec:prop} presents the proposed method of learned lossless image compression through interpolation. Section \ref{sec:expres} presents experimental results and comparisons and Section \ref{sec:con} concludes the paper.



\section{Scale-based auto-regressive models}\label{sec:relwrk}

In L3C \cite{mentzer2019practical}, which is the first to propose a scale-based auto-regressive model, a three scale model is used. First, all scale representations denoted $z^{(i)}, ~ i=1,2,3$ are obtained at the encoder by processing the previous scale representation with encoder neural networks $E^{(i)}(.)$ and then quantizing the scale representation variables to the nearest integer:
\begin{align}
    x^{(i+1)} & \leftarrow E^{(i)}(x^{(i)}), \nonumber \\
    z^{(i+1)} & \leftarrow Q(x^{(i+1)}) \qquad i=0,1,2    
\end{align}
The original image is the initial scale representation $x^{(0)}$ and $Q(.)$ is the quantization operator. Each scale representation has half the horizontal and vertical resolution of the previous scale. The last scale's probability distribution is assumed uniform and each preceding scale's probability distribution is conditioned on the next scale:
\begin{align}
    & p(z^{(3)}) \sim uniform \\
    & p(z^{(i-1)} | z^{(i)})  \qquad i=3,2 \\
    & p(x^{(0)} | z^{(1)})
\end{align}
Based on the probability models, first, the last scale $z^{(3)}$ is transmitted by the encoder to the decoder with simple fixed length codes. Then, starting with $z^{(3)}$, both the encoder and decoder process the scale representation $z^{(i)}$ with neural networks to obtain the conditional probability distribution of the previous scale $p(z^{(i-1)} | z^{(i)})$ and encode/decode it. This is repeated until the original image is encoded/decoded.

The lossless image compression through super resolution (SReC) \cite{cao2020lossless} paper has a framework similar to L3C but the major differences are how the multiscale representations are obtained and the progressive modeling of the probability distribution of each scale. Unlike in L3C, the scale representations $x^{(i)}$ are obtained simply by average pooling every 2x2 pixel group of the previous scale and rounding to the nearest integer:
\begin{align}
    x^{(i+1)} & \leftarrow AvgPool_{2x2}(x^{(i)}) \qquad i=0,1,2  
\end{align}
Each scale representation's pixels are split to four groups based on even and odd rows and columns:
\begin{align}
    \{ x^{(i)}_{00}, x^{(i)}_{01}, x^{(i)}_{10}, x^{(i)}_{11} \} & \leftarrow x^{(i)} 
\end{align}

Starting with the second last scale ($i=3$), 
each scale's probability distribution is progressively obtained by modeling each group's probability distribution conditioned on the next scale and the previous groups in the same scale:
\begin{align}
    & p(x^{(i-1)}_{00} | x^{(i)}),  \label{eq:srec_prob} \\
    & p(x^{(i-1)}_{01} | x^{(i)}, x^{(i-1)}_{00}),  \\
    & p(x^{(i-1)}_{10} | x^{(i)}, x^{(i-1)}_{00}, x^{(i-1)}_{01}),  
\end{align}
The last group's pixels ($x^{(i-1)}_{11}$) are obtained from the first three groups and the next scale since the next scale was obtained by average pooling: 
\begin{align}
     x^{(i-1)}_{11} &\leftarrow 4x^{(i)} - x^{(i-1)}_{00} - x^{(i-1)}_{01} - x^{(i-1)}_{10} + \{-1/0/1/2\},  
\end{align}
Note that some additional bits (represented by $\{-1/0/1/2\}$ in the above equation) are transmitted by the encoder to accurately find the $x^{(i-1)}_{11}$ pixels due to the rounding operation in the average pooling. Finally, all groups can then be combined to form the scale's representation:
\begin{align}
     x^{(i-1)}  &\leftarrow \{ x^{(i-1)}_{00}, x^{(i-1)}_{01}, x^{(i-1)}_{10}, x^{(i-1)}_{11} \}   
     \label{eq:srec_comb}
\end{align}
Equations (\ref{eq:srec_prob})-(\ref{eq:srec_comb}) are repeated for the next scales ($i=2,1$) to complete the probability model of the entire image.

Based on the probability models, first, the last scale $x^{(3)}$ is transmitted by the encoder to the decoder with simple fixed length codes. Then, both the encoder and decoder process the scale representation $x^{(3)}$ with neural networks to obtain the conditional probability distribution of the initial group in the previous scale  $p(x^{(i-1)}_{00} | x^{(i)})$ and encode/decode it. Similar processing with neural networks is performed to obtain the conditional probability distribution of the other groups based on Equation (8) and (9) and they are  encoded/decoded. Then Equation (10) and (11) are applied to obtain the previous scale representation $x^{(i-1)}$. Then the processing starts again with Equation (7) for the new scale. This is repeated until the original image $x^{(0)}$ is encoded/decoded.

In MSPSM \cite{zhang2020lossless}, the overall framework is similar to that of SReC but the scale representations $x^{(i)}$ are obtained by not average pooling but simply splitting the original image into 4 groups based on even and rows and columns or 6 groups for improved performance. Then the initial group is split again to form the next scale. Starting with the initial group in the last scale, each group's probability distribution is modeled conditioned on the previous groups in a progressive manner. Note that since groups are obtained by splitting, once all groups in a scale are decoded, the previous scale's initial group is readily available.

The compression performance of L3C is better than all classical lossless compression systems except FLIF \cite{sneyers2016flif}. SReC improves the compression performance over L3C significantly and outperforms also FLIF. MSPSM presents three models with increasing complexity and the compression performance of the first one is similar to that of SReC while the latter two have better performance. While L3C, SReC and MSPSM provide state of the art lossless compression performance and reasonable encoding/decoding speeds on GPUs they require neural networks with millions of parameters (L3C: 5.0M, SReC: 4.2M, MSPSM:1.9M/9.9M) and high computational complexity. This paper presents a scale-based auto-regressive system that has similarities to SReC and MSPSM but several modifications in the overall system and neural network design are proposed, which all help to improve compression performance and/or reduce complexity, to achieve the same compression performance as SReC and the first model in MSPSM with 10x (or more) less parameters and computational complexity.

\section{Learned Lossless Image Compression Trough Interpolation (LLICTI)} \label{sec:prop}
This section presents the proposed method of learned lossless image compression through interpolation (LLICTI) in four sub-sections. 

\subsection{Overall System Architecture and Decoding Procedure} \label{ssec:sys}

\begin{figure}[tbh]
	\centering
	\includegraphics[width=\linewidth]{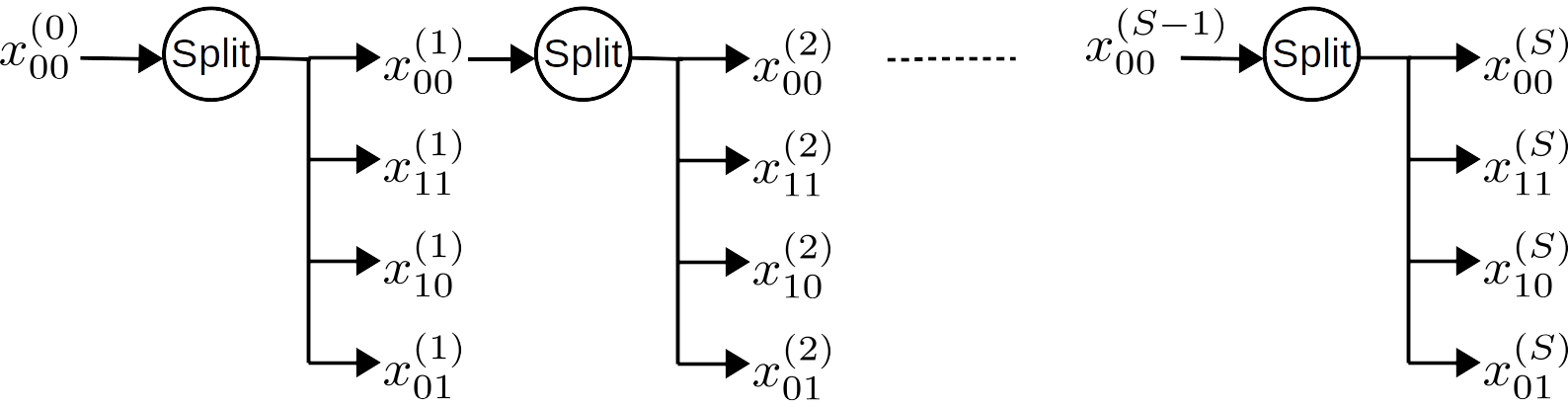} 
	\caption{Multi-scale representation of original image $x_{00}^{(0)}$ in LLICTI.}
	\label{fig:mulscl}
\end{figure}

A multi-scale representation of the original input image is obtained at the encoder by first splitting it into 4 subbands based on even and odd indices of the rows and columns. This splitting is then repeated on the first subband of each scale to obtain a multi-scale representation of the original image. The obtained multi-scale representation with subbands $x_{mn}^{(i)}$, where $i$ denotes the scale index and $(m,n)\in \{(0,0), (0,1), (1,0), (1,1)\}$ denote the subband index, can be represented as follows if the original image is denoted $x_{00}^{(0)}$ and there are $S$ scales (see also Figure \ref{fig:mulscl}):
\begin{align}
    \{ x^{(i+1)}_{00}, x^{(i+1)}_{01}, x^{(i+1)}_{10}, x^{(i+1)}_{11} \} & \leftarrow x_{00}^{(i)} \qquad i=0,...,S-1 
\end{align}

The decoding procedure starts with the last scale by decoding the $x^{(S)}_{00}$ subband with fixed length codes of their pixel values in RGB color space. The pixels decoded with fixed length codes is $\frac{1}{4^{S}}$ of all original image pixels and is a negligible fraction for large $S$, such as $S=5$ which we use in our experiments.

\begin{figure}[tbh]
	\centering
	\includegraphics[width=1.0\linewidth]{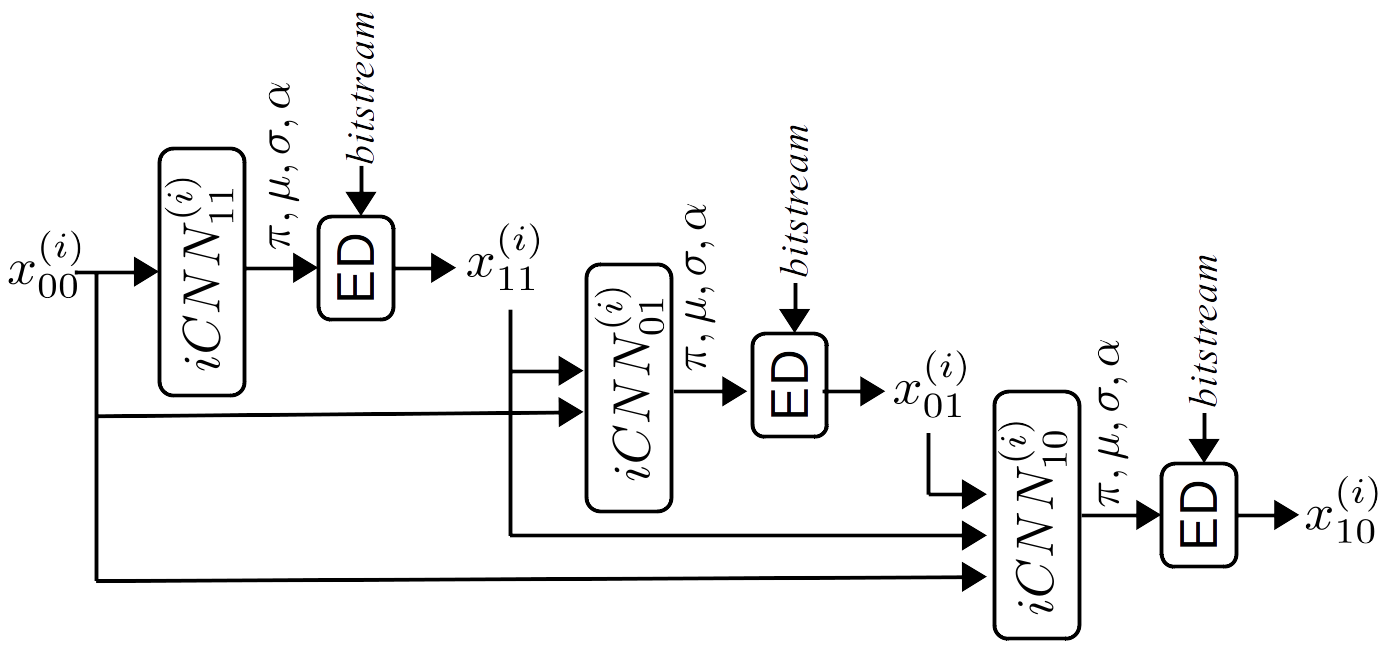} 
	\caption{Decoding procedure for one scale. Three interpolator convolutional neural networks ($iCNN$) are used in each scale.}
	\label{fig:dec}
\end{figure}

The processing steps to decode the remaining subbands in this scale are summarized in Figure \ref{fig:dec}. First, the $x^{(S)}_{00}$ subband is processed at the decoder with an $\boldsymbol{i}$nterpolator $\boldsymbol{c}$onvolutional $\boldsymbol{n}$eural $\boldsymbol{n}$etwork ($iCNN^{(S)}_{11}$) to obtain the probability distribution parameters of the next subband $x^{(S)}_{11}$, which is then decoded by an entropy decoder from the bitstream. Next, $x^{(S)}_{00}$ and $x^{(S)}_{11}$ subbands are the inputs to another interpolator convolutional neural network ($iCNN^{(S)}_{01}$) to obtain the probability distribution parameters of the $x^{(S)}_{01}$ subband, which is then decoded. Then, in a similar manner, the $x^{(S)}_{10}$ subband is decoded. Finally, all the decoded subbands are combined to obtain the initial subband of the next scale to decode:
\begin{align}
    x_{00}^{(i-1)} \leftarrow \{ x^{(i)}_{00}, x^{(i)}_{01}, x^{(i)}_{10}, x^{(i)}_{11} \}  \qquad i=S 
\end{align}
To decode the next scales, the above described processing steps are repeated for each scale ($i=S-1,...,1$), yielding the decoding of the original image $x_{00}^{(0)}$. The encoder performs similar and appropriate processing to encode the image.

Note that the described decoding procedure corresponds to an interpolation based scheme which starts with an image grid of every $(2^S)^{th}$ pixel (i.e. $x^{(S)}_{00}$) and interpolates progressively missing pixels of the grid populating the entire image, thus the name of the method.

\subsection{Probability Model} \label{ssec:prob}
The probability model of an original image, upon which the system architecture and decoding procedure in Section \ref{ssec:sys} is based on, can be described as follows. The pixels in the initial subband of the last scale $x^{(S)}_{00}$ are assumed to follow a uniform distribution:
\begin{align}
    p(x^{(S)}_{00})~ \sim ~ & uniform 
    \label{eq:unif}
\end{align}
The probability distribution of the initial subband of the preceding scale $p(x^{(S-1)}_{00})$ is then factorized as the product of conditional distributions of each subband of the scale conditioned progressively on preceding subbands:
\begin{align}
    p(x^{(i-1)}_{00}) = ~ & p(x^{(i)}_{00})\times \nonumber \\
                        ~ & p(x^{(i)}_{11}~|~x^{(i)}_{00})\times \nonumber \\
                        ~ & p(x^{(i)}_{10}~|~x^{(i)}_{00},x^{(i)}_{11})\times \nonumber \\
                        ~ & p(x^{(i)}_{01}~|~x^{(i)}_{00},x^{(i)}_{11},x^{(i)}_{10}), \qquad i=S 
    \label{eq:facsub}
\end{align}
Note that Equation (\ref{eq:facsub}) can be applied recursively for each scale ($i=S-1,...,1$) to factorizate the distribution of the original image $x^{(0)}_{00}$ into the conditional distributions of subbands, conditioned on previous subbands, with an initial marginal distribution given in Equation (\ref{eq:unif}).

To describe each conditional distribution in Equation (\ref{eq:facsub}), let $Y^{(i)}_{mn}$ denote the collection of subbands that $x^{(i)}_{mn}$ conditions on. Further, let  $x^{(i)}_{mn}[u,v]$ denote the pixel at spatial location $(u,v)$ and $C^{(i)}_{mn}[u,v]$ denote the collection of pixels in $Y^{(i)}_{mn}$ that the pixel $x^{(i)}_{mn}[u,v]$ conditions on through the receptive field of the interpolator CNN ($iCNN^{(i)}_{mn}$). Then each conditional distribution in Equation (\ref{eq:facsub}) is factorized as follows:
\begin{align}
    p(x^{(i)}_{mn}~|~Y^{(i)}_{mn}) = \prod_{u,v} p(&x^{(i)}_{mn}[u,v]~|~C^{(i)}_{mn}[u,v]), \nonumber \\
                                                 & (m,n)\in \{(1,1),(1,0),(0,1)\}
    \label{eq:facpix}
\end{align}

To describe the conditional distributions of each pixel in Equation (\ref{eq:facpix}), let us simplify the notation by dropping the spatial location, subband and scale indices and simply denote a pixel by $x$ and the collection of pixels it is conditioned on by $C$. Then, the conditional distribution of each pixel in Equation (\ref{eq:facpix}) can be further factorized over distributions of its RGB color components ($x=\{x^{(r)}, x^{(g)}, x^{(b)}\}$), i.e. sub-pixels\footnote{Here, we follow the convention in the related previous research and use sub-pixel to denote each color component and pixel to denote all color components together}, as follows:
\begin{align}
    p(x^{(r)}, x^{(g)}, x^{(b)}|C) = ~ & p(x^{(r)}| C) \times \nonumber \\ 
                                        & p(x^{(g)}| C, x^{(r)}) \times ~ \nonumber \\ 
                                        & p(x^{(b)}| C, x^{(g)}, x^{(b)}) 
\label{eq:clr}
\end{align}

Each conditional distribution in Equation (\ref{eq:clr}) is modeled with a discretized Gaussian mixture model (GMM) given below in Equation (\ref{eq:gmm}), where $F(.)$ denotes the cumulative distribution function (CDF) of the standard Gaussian distribution and $\mu^{(c)}_{i}$, $\sigma^{(c)}_{i}$ and $\pi^{(c)}_i$ are the mean, standard deviation and the weight of the $i^{th}$ mixture of the GMM, respectively, of sub-pixel $x^{(c)} \in \{x^{(r)}, x^{(g)}, x^{(b)}\}$:
\begin{align}
p(x^{(c)};~{{\pi^{(c)},\mu^{(c)},\sigma^{(c)}}}) = \sum _{i=1}^{K} \pi^{(c)}_{i}[&F(\frac{x^{(c)}+0.5-\mu^{(c)}_{i}}{\sigma^{(c)}_{i}}) \nonumber \\
-&F(\frac{x^{(c)}-0.5-\mu^{(c)}_{i}}{\sigma^{(c)}_{i}})]
\label{eq:gmm} 
\end{align}

The GMM parameters $\mu^{(c)}_{i}$, $\sigma^{(c)}_{i}$ and $\pi^{(c)}_i$ are produced by the respective interpolator CNN processing the conditioning pixels $C$. The conditioning of each sub-pixel on also the preceding sub-pixels in Equation (\ref{eq:clr}) is accomplished with the simple but efficient method of Salimans in PixelCNN++ \cite{salimans2017pixelcnn++}, which simply updates the means by multiplication of preceding sub-pixels with scaling coefficients ($\{a,b,c\}=\alpha$) that are produced also by the interpolator convolutional neural network:
\begin{align}
\mu^{(r)} = & \mu^{(r)}(C)  \nonumber \\
\mu^{(g)} = & \mu^{(g)}(C) +a(C)x^{(r)} \nonumber \\
\mu^{(b)} = & \mu_{(b)}(C) +b(C)x^{(r)}+ c(C)x^{(g)}   
\label{eq:mnup_b}
\end{align}

\subsection{General Neural Network Architecture of One Interpolator} \label{ssec:intr}
\begin{figure}[tbh]
	\centering
	\includegraphics[width=0.95\linewidth]{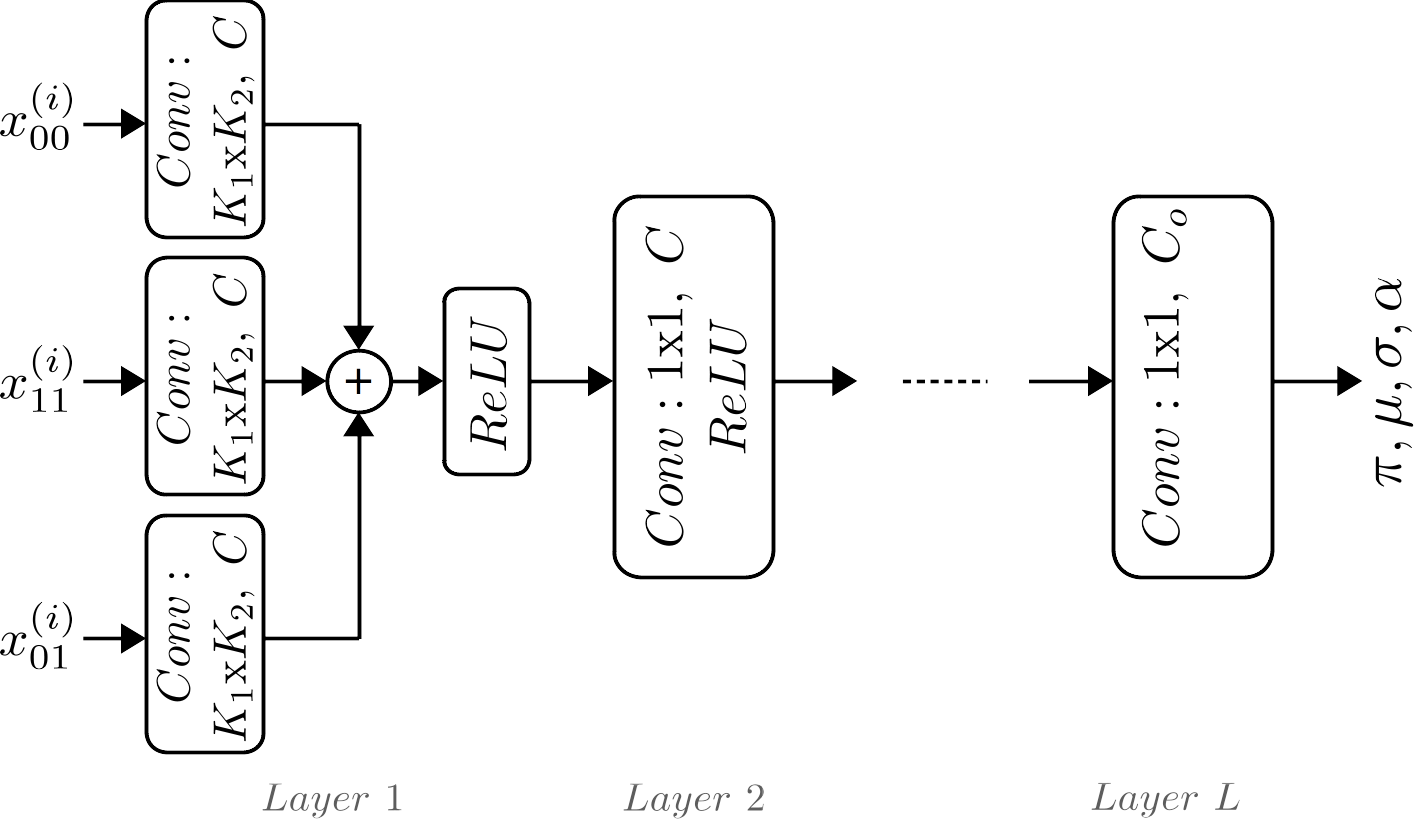}
	\caption{Architecture of interpolator Convolutional Neural Network at scale $i$ for sub-band $x_{10}$ ($iCNN_{10}^{(i)}$). The architectures for the other two interpolators of a scale,  $iCNN_{01}^{(i)}$ and $iCNN_{11}^{(i)}$, are similar and lack the bottom and bottom two convolutional layers in Layer 1, respectively.}
	\label{fig:intnn}
\end{figure}

The neural network architecture used for all interpolators $iCNN^{(i)}_{mn}$, $i=1,...,S$ and $(m,n)\in \{(1,1),(1,0),(0,1)\}$ in the proposed LLICTI system is shown in Figure \ref{fig:intnn}. It has $L$ layers, where layer 2 to $L$ comprise convolution layers with 1x1 kernel size, stride of 1 and (except the last layer) ReLU activation functions. The first layer contains one, two or three convolutional layers, depending on how many previous subbands $x_{mn}^{(i)}$ are inputs to this interpolator (see Figure \ref{fig:dec}). The outputs are added and processed with ReLU activation.

\begin{figure}[tbh]
	\centering
	\includegraphics[width=1.0\linewidth]{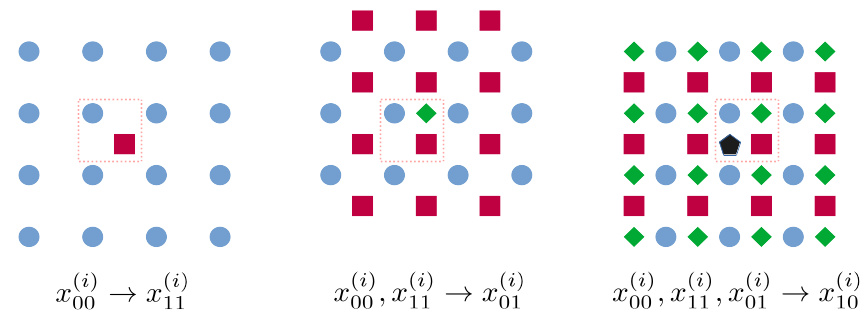} 
	\caption{Interpolator layer 1 kernel sizes. $x_{00}^{(i)}(4\text{x}4) \rightarrow x_{11}^{(i)}$
        $x_{00}^{(i)}(3\text{x}4), x_{11}^{(i)}(4\text{x}3) \rightarrow x_{01}^{(i)}$ $x_{00}^{(i)}(4\text{x}3), x_{11}^{(i)}(3\text{x}4), x_{01}^{(i)}(4\text{x}4) \rightarrow x_{10}^{(i)}$
    }
	\label{fig:grid}
\end{figure}

Since all layers, except the first, have filters with kernel size 1x1, the receptive field of the $iCNN$ or the pixels of the previous subbands used for interpolation are determined by the kernel sizes of the first layer convolutions. Due to the relative positions of different subbands on the sampling grid, the kernel sizes were determined based on the input subband and the subband to be interpolated, as shown in Figure \ref{fig:grid}.

To train the parameters of all interpolator convolutional neural networks, the cross entropy loss function 
\begin{equation}
    \mathbb{E}_{q(x_{00}^{(0)})}[-\log p(x_{00}^{(0)})] 
\end{equation}
is used, where $q(.)$ is the true distribution of the image and $p(.)$ is the distribution of the image obtained by the presented probability model in Equations (\ref{eq:unif})-(\ref{eq:mnup_b}). In practice, the loss simply becomes the sum of all $-\log p(.)$ quantities of all interpolated pixels in the training batch where $p(.)$ are the probability distributions provided by the interpolators for each interpolated pixel.

\subsection{System and NN Architecture Parameters to Analyze} \label{ssec:anal}
The proposed LLICTI system described so far has 3 interpolators ($iCNN$) per scale, for a total of $3S$ interpolators overall, where $S$ is the total number of scales. There are many parameters of the overall system and the $iCNN$ architecture that we will analyze. In Section \ref{ssec:expana}, we examine
\begin{itemize}
    \item whether compressing pixels in the RGB or YCoCg-R color space,
    \item whether obtaining probability distribution parameters $\pi,\mu,\sigma,\alpha$ using a single or using multiple independent interpolator CNNs
    \item whether using the same interpolator CNN (i.e. with same weights) in several scales
\end{itemize}
performs better in terms of compression-complexity trade-off. We also examine standard neural network architecture parameters such as, the number of layers and the number of channels of each layer for better compression-complexity trade-off.

\section{Experiments and Results}\label{sec:expres}
This section first presents experimental settings and then provides experimental analyses and results of the proposed learned lossless image compression through interpolation (LLICTI) systems. Comparisons with the literature in terms of compression performance, number of used parameters, encoding/decoding times and computational complexity estimates are also provided.

\subsection{Experimental Settings} \label{ssec:expset}
All LLICTI systems to be discussed in Section \ref{ssec:expana} are trained with the Open Images training dataset prepared by the authors of L3C \cite{mentzer2019practical}. The Adam optimizer \cite{kingma2014adam} is used for optimization with a batch size of 64. The learning rate is initialized to $10^{-4}$ and is reduced with a decay factor of $0.5$ (up to a minimum of $10^{-5}$) when the validation cost plateaus. The training takes typically around $500K$ weight updates. 

The LLICTI systems are tested with the Open Images test dataset prepared again by the authors of L3C \cite{mentzer2019practical}, which includes 500 images. The arithmetic coder provided by \cite{mentzer2019practical}, which runs on the CPU \cite{torchac_github}, is used for encoding to and decoding from a bitstream during the tests. Note that all neural network operations are performed on the GPU and the obtained probability tables are copied to the CPU memory and the arithmetic encoding/decoding operations are performed on the CPU in our tests of LLICTI. 
PyTorch \cite{paszke2019pytorch} is used to implement both training and testing. Our codes are available on GitHub \cite{llicti_github}.

\subsection{Experimental Analysis and Ablation Study} \label{ssec:expana}
The following experiments summarized in Table \ref{tb:anal} are conducted to analyze several parameters of the overall compression system and the interpolator CNN architecture, as discussed in Section \ref{ssec:anal}. Note that all interpolators ($iCNN^{(i)}_{mn}$ $i=1,...,S$ and $(m,n)\in \{(1,1),(1,0),(0,1)\}$) can be trained independently from each other since their inputs can be obtained directly from the original image and their outputs contribute to the cost (i.e. total bitrate) independently. Considering also the fact that compression of scale 1 subbands $x_{mn}^{(1)}$, $(m,n)\in \{(1,1),(1,0),(0,1)\}$ is more important (as they comprise 4 times as many pixels as scale 2, which comprises 4 times as many pixels as scale 3 etc) than that of other scale subbands, the initial experiments discussed below are conducted by compressing only scale 1 subbands (i.e. $x_{00}^{(1)}$ is assumed available and  $x_{11}^{(1)}, x_{01}^{(1)}, x_{10}^{(1)}$ are compressed). 

\newcommand{\ra}[1]{${#1}\rightarrow$} 
\newcommand{\ua}{$\uparrow$} 

\begin{table*}[tbh]
\setlength\tabcolsep{4.5pt} 
\centering
\caption{\textsc{Analysis of system and neural network architecture parameters}}
\label{tb:anal}
\begin{tabular}{ ll|c|c|c|c|c|c|c|c|c|c|c|c|c|c } 
\hline
Experiment \#                             & \hspace{-0.4cm}        & 1     & 2     & 3     & 4     & 5     & 6     & 7     & 8     & 9     & 10     & 11    & 12     & 13     & 14     \\ 
(Reference Experiment)                    & \hspace{-0.4cm}        &       & (1)   & (2)   & (3)   & (4)   & (5)   & (4)   & (7)   & (8)   & (9)    & (10)  & (11)   & (12)   & (13)   \\ \hline
Color space: RGB/YCoCg-R                  & \hspace{-0.4cm}        & RGB   & YCC   & YCC   & YCC   & YCC   & YCC   & YCC   & YCC   & YCC   & YCC    & YCC   & YCC    & YCC    & YCC    \\ \hline
($\pi,\mu,\sigma,\alpha$): Joint/Separate & \hspace{-0.4cm}        & JNT   & JNT   & SEP   & SEP   & SEP   & SEP   & SEP   & SEP   & SEP   & SEP    & SEP   & SEP    & SEP    & SEP    \\ \hline
\# Layers in $iCNN$                       & \hspace{-0.4cm}        & 4     & 4     & 4     & 3     & 2     & 2     & 3     & 3     & 3     & 3      & 3     & 3      & 3      & 3      \\ \hline
\multirow{5}{*}{\# Channels}              & \hspace{-1.2cm}Scale 1 & 96    & 96    & 44    & 44    & 44    & 64    & 60    & -     & -     & -      & -     & -      & 60     & 88     \\ 
                                          & \hspace{-1.2cm}Scale 2 & -     & -     & -     & -     & -     & -     & -     & 44    & 32    & -      & -     & -      & \ua    & \ua    \\ 
                                          & \hspace{-1.2cm}Scale 3 & -     & -     & -     & -     & -     & -     & -     & -     & -     & 36     & 24    & 24     & -      & \ua    \\ 
                                          & \hspace{-1.2cm}Scale 4 & -     & -     & -     & -     & -     & -     & -     & -     & -     & 30     & 20    &\ua     & -      & \ua    \\ 
                                          & \hspace{-1.2cm}Scale 5 & -     & -     & -     & -     & -     & -     & -     & -     & -     & 30     & 20    &\ua     & -      & \ua    \\ \hline 
Share $iCNN$ across scales                & \hspace{-0.4cm}        & -     & -     & -     & -     & -     & -     & -     & -     & -     & F      & F     & T      & T      & T      \\ \hline \hline
\multirow{5}{*}{Compression (bpp)}        & \hspace{-1.2cm}Scale 1 & 5.781 & 5.736 & 5.607 & 5.605 & 5.658 & 5.624 & 5.524 & -     & -     & -      & -     & -      & 5.525  & 5.465  \\ 
                                          & \hspace{-1.2cm}Scale 2 & -     & -     & -     & -     & -     & -     & -     & 1.852 & 1.862 & -      & -     & -      & 1.868  & 1.841  \\ 
                                          & \hspace{-1.2cm}Scale 3 & -     & -     & -     & -     & -     & -     & -     & -     & -     & 0.551  & 0.556 & 0.556  & -      & 0.553  \\ 
                                          & \hspace{-1.2cm}Scale 4 & -     & -     & -     & -     & -     & -     & -     & -     & -     & 0.158  & 0.159 & 0.158  & -      & 0.160  \\
                                          & \hspace{-1.2cm}Scale 5 & -     & -     & -     & -     & -     & -     & -     & -     & -     & 0.044  & 0.044 & 0.044  & -      & 0.045  \\ \hline
Total \# Parameters ($\times 10^3$)       & \hspace{-0.4cm}        & 93    & 93    & 94    & 72    & 49    & 72    & 109   & 72    & 48    & 144    & 88    & 34     & 109    & 188    \\ \hline
\end{tabular}
\end{table*}

Note that a primary goal of the LLICTI method is to obtain a lossless image compression system that has significantly less complexity than other similar systems in the literature while achieving the best possible compression performance. In particular, the goal is to have at least 10x less parameters in the LLICTI system than the smallest neural network in the related literature, which is the normal model in MSPSM with $1.8$M parameters. The parameters, in particular number of layers and channels, of the LLICTI systems in the following experiments were chosen to chase that goal. 

The first conducted experiment is experiment 1, which can be considered the default system and compresses pixels in the RGB color space, uses a single (i.e. joint) interpolator CNN to obtain all probability distribution parameters ($\pi,\mu,\sigma,\alpha$), has $L=4$ layers in the interpolator CNNs and has $C=96$ channels in each convolutional layer. This system achieves a compression performance of 5.781 bits per pixel (bpp) and requires 93K parameters. Note that the 5.781 bpp is obtained by dividing the total number of bits used for encoding $x_{11}^{(1)}, x_{01}^{(1)}$ and $x_{10}^{(1)}$ by the total number of pixels in the original image, and the bits required for compressing $x_{00}^{(1)}$ are not considered in this and following similar experiments.

Next, it is investigated whether/how the color space of input pixels can improve compression-complexity trade-off and experiment 2 is conducted, where the original image is converted to the YCoCg-R color space \cite{malvar2003ycocg}, which is integer-to-integer invertible, and then compressed. While the interpolator CNNs can learn the dependencies in the RGB color space and account for them, it may be beneficial in terms of compression-complexity trade-off to provide the input to the neural networks in a decorrelated color space and allocate neural network capacity to learn other dependencies. The system in experiment 2 achieves an improved compression performance of 5.736 bpp and requires the same 93K parameters. Hence, YCoCg-R color space is used in the next experiments.

Next, it is investigated whether each interpolator CNN should comprise a single CNN that produces all parameters ($\pi,\mu,\sigma, \alpha$) required for the probability distribution model or whether each interpolator should comprise 4 separate CNNs, each producing one of the parameters. Since each of these parameters affect the probability distribution (and thus the cost function) in different ways, learning them trough separate neural networks could be better in terms of compression-complexity trade-off. Hence, experiment 3 is conducted, which differs from its reference (experiment 2), by the use of 4 separate CNNs for each interpolator and the reduced number of channels (from 96 to 44) in each CNN to keep the total number of parameters similar. The system in experiment 3 achieves an improved compression performance of 5.607 bpp and requires a similar 94K parameters. Hence, in the next experiments, for each interpolator, 4 separate CNNs are used, each of which produces one of the parameters ($\pi,\mu,\sigma,\alpha$).

Next, the number of layers and channels in the interpolator CNNs are examined in experiments 4-7. In experiment 4, the number of layers in the interpolator CNNs are reduced from 4 to 3 and almost the same compression performance of 5.605 bpp as reference experiment 3 is achieved, while the required number of parameters drops from 94K to 72K. Based on this result, the number of layers in the interpolator CNNs are further reduced to 2 in experiment 5, however, the compression performance deteriorated to 5.658 bpp while 49K parameters were required. Thus experiment 6 is conducted, where the number of layers is kept at 2 but the number of channels is increased to 64 to achieve the same number of parameters (72K) as experiment 4. The achieved compression performance is 5.624 bpp, which is inferior to the result of experiment 4. Hence, experiment 7 is conducted, which keeps the number of layers at 3 and increases the number of channels to 60 to achieve a compression performance of 5.524 bpp with 109K required parameters.

It is now assumed that several parameters of the overall system and the interpolator CNN architecture have been identified for better compression-complexity trade-off. In particular, compressing pixels in the YCoCg-R space, using 4 separate CNNs for each interpolator to obtain parameters ($\pi,\mu,\sigma,\alpha$) and using 3 layers in the interpolator CNNs have been identified to provide better compression-complexity trade-off and will be used in the next experiments.

Next, in experiments 8 and 9, compressing scale 2 subbands (i.e. $x_{00}^{(2)}$ is assumed available and  $x_{11}^{(2)}, x_{01}^{(2)}, x_{10}^{(2)}$ are compressed) is considered and only the number of channels in the interpolator CNNs are examined. 
In experiment 8, 44 channels are used to achieve a compression performance of 1.852 bpp with 72K required parameters. In experiment 9, the number of channels is reduced to 32 and a slightly worse compression performance of 1.862 bpp is obtained with 48K required parameters. 

In experiments 10 and 11, scale 3, 4, and 5 subbands are compressed. In experiment 10, 36, 30 and 30 channels are used to achieve a compression performance of 0.551 bpp, 0.158 bpp and 0.044 bpp for scales 3, 4 and 5, respectively, with a total 144K required parameters. In experiment 11, the number of channels is reduced to 24, 20 and 20 for scales 3, 4 and 5, respectively, and the obtained compression performance is slightly inferior but the required total number of parameters drops by a significant amount to 88K.

Next, it is investigated whether using the same interpolator CNN (i.e. with same weights) in several scales is beneficial for compression-complexity trade-off. Although the dependency of pixels in different scales may be different, it is likely that there are significant amount of common features and sharing interpolator CNNs across scales may improve the compression-complexity trade-off. Hence, in experiment 12, the interpolator CNNs for scale 3 with 24 channels are shared across scales 3, 4 and 5 (Note that they are trained from scratch using the total cost of the three scales.) The compression performance is almost identical to that of the reference experiment 11 for all scales but the total number of required parameters drops significantly from 88K to 34K. This is a surprising and significant result. Inspired by this result, experiment 13 is conducted, where the interpolator CNNs (with the same 60 channels as in experiment 7) are shared across scales 1 and 2. The obtained compression performance of 5.525 bpp for scale 1 is identical to that of experiment 7 and scale 2 is compressed at a competitive 1.868 bpp (see experiments 8, 9). Hence, sharing interpolator CNNs across scale 1 and 2 proved to be also very beneficial in terms of compression-complexity trade-off.

Finally, experiment 14 is conducted, where the same interpolator CNNs are shared across all scales 1-5 and the number of channels in the convolution layers is set to 88 to achieve a total 188K parameters, which is 10x less (our initial goal) than the smallest neural network in the literature \cite{zhang2020lossless}. The compression performance is the best so far in the experiments for scales 1 and 2, achieving 5.465 bpp and 1.841 bpp, respectively. The compression performance for the remaining scales 3-5 is also very competitive as can be seen in Table \ref{tb:anal}. The total bitrate for all scales becomes 8.064 bpp (which dos not include the 0.023 bpsp bitrate of $x_{00}^{(5)}$ that is to be coded with fixed-length codes of 8 bits per symbol.) Hence, the system and the neural network architecture for the interpolators in experiment 14 is chosen to be the best in terms of compression-complexity trade-off and is compared to other systems in the literature in the next section.

\subsection{Comparisons with Related Literature} \label{ssec:comp}
This section compares the proposed LLICTI to other lossless image compression systems in the literature in terms of compression performance, number of required parameters, encoding/decoding times and computational complexity. The comparison results are summarized in Table \ref{tb:compar}. (Note that in these results of the LLICTI system, all necessary information is written into the bitstream including $x_{00}^{(5)}$ and control information, such as the number of scales, height and width of every scale and the image. The LLICTI system can also compress arbitrary sized images.)

\begin{table}[tbh]
\setlength\tabcolsep{1.5pt} 
\centering
\caption{\textsc{Comparisons with Related Literature}}
\label{tb:compar}
\begin{tabular}{ l|c|c|c|c|c } 
\hline
 & Method & bpsp(bpp) & Param- & Enc/Dec    & Enc/Dec \\
 &        &      &      eters  & Computat.  & Time    \\
 &        &      &             & (KMAC/pix) & (sec)   \\
\hline
\multirow{4}{4.2em}{Traditional Methods} 
& PNG      & 4.01(12.03) & - & - & -  \\ 
& JPEG2000 & 3.06(9.18) & - & - & -  \\
& WebP     & 3.05(9.15) & - & - & -  \\
& FLIF \cite{sneyers2016flif} & 2.87(8.61) & - & - & -  \\
\hline
\multirow{5}{4.2em}{Learning Based Methods} 
& L3C \cite{mentzer2019practical}       & 2.99(8.97) & 5.01 M  & 686/428 & 0.67/0.61 \\ 
& SReC \cite{cao2020lossless}           & 2.70(8.10) & 4.20 M  & 878/878 & 0.56/\textbf{0.59} \\
& MSPSM(norm)\cite{zhang2020lossless}  & 2.71(8.14) & 1.87 M & -         & -       \\
& MSPSM(big) \cite{zhang2020lossless}   & \textbf{2.63(7.88)} & 1.87 M & -         & -       \\
& \textbf{LLICTI (Ours)}                & 2.70(8.10) & \textbf{0.19} M & \textbf{66/66} & \textbf{0.53}/0.70    \\
\hline
\end{tabular}
\end{table}

\subsubsection{Compression performance and number of parameters} \label{ssec:com}
The compression performance results for the Open Images test dataset \cite{mentzer2019practical} are given in terms of bits per sub-pixel (bpsp) and bits per pixel (bpp). The best performing traditional method is FLIF \cite{sneyers2016flif} and it achieves an average compression performance of 2.87 bpsp. Among the learning based methods, the first proposed scale-based auto-regressive method L3C \cite{mentzer2019practical} achieves a compression performance of 2.99 bpsp. Next, SReC \cite{cao2020lossless}, MSPSM(normal) \cite{zhang2020lossless} and the proposed LLICTI achieve a compression performance of 2.70 bpsp. A common feature of these three scale-based auto-regressive methods is that they perform the compression/probability-modeling of each scale in 3 progressive steps. MSPSM(big) achieves an improved compression performance of 2.63 bpsp as it performs the compression/probability-modeling of each scale in 6 progressive steps \cite{zhang2020lossless}.

If the learned systems are compared in terms of number of parameters of the neural networks that they use, it can be seen that the proposed LLICTI method uses 0.19 M (188K) parameters, which is about 10x less than MSPSM, 22x less than SReC and 26x less than L3C. Despite the significantly less number of parameters of LLICTI, compression performance similar to these methods was achieved. The improved compression-complexity trade-off of LLICTI is due to several factors, such as simpler neural network design, sharing the same interpolator neural networks across different scales, using separate neural networks for different parameters of the probability distribution and performing the compression in the YCoCg-R color space instead of the RGB color space.

\subsubsection{Computational complexity and encoding/decoding times} \label{ssec:cc}
The computational complexity of the neural network calculations for encoding and decoding in the learning based systems are compared using the multiply-accumulate (MAC) operations counter tool in \cite{flops}. As shown in Table \ref{tb:compar}, the proposed LLICTI system requires more than 10x less MACs than L3C and SReC for both encoding and decoding, which is mainly due to the less number of parameters of the used neural networks. (Note that MSPSM is not included in computation or encoding/decoding times comparisons since their codes are not shared.)

The average encoding and decoding time of an image (averaged over all 500 images on the Open Images test dataset where the average image resolution is $\sim$768x576) measured on our computing system with an NVIDIA GeForce RTX 2080 GPU and Intel i7-9700 CPU are also given in Table \ref{tb:compar}. L3C, SReC and the proposed LLICTI achieve similar average encoding/decoding times under one second and close to half a second. However, the implementations of L3C and SReC enjoy an advantage over that of LLICTI. In particular, implementations of L3C and SReC don't copy the calculated CDFs (necessary for arithmetic coding on the CPU) from GPU to CPU but use a custom CUDA kernel to store the CDFs 
in managed/unified memory accessible from both CPU and GPU (see L3C appendix \cite{mentzer2019practical} and GitHub page \cite{l3c_github}.) On the other hand, LLICTI implementation moves the CDF tensors from GPU to CPU for arithmetic coding. If the LLICTI implementation would also store CDF tensors in unified/managed memory, its encoding/decoding times are likely to decrease. Lastly, while encoding/decoding times of MSPSM could not be obtained on our computing system (since their codes are not shared), it is reported in their paper \cite{zhang2020lossless} that their encoding/decoding times are about $20\%$ and $100\%$ longer than those of SReC, for their normal and big systems, respectively.

\section{Conclusions} \label{sec:con}
This paper presented a learned lossless image compression method based on a progressive interpolation scheme. First, a subset of pixels of the original image grid (e.g. every $32^{nd}$ pixel horizontally and vertically) is encoded/decoded using simple fixed-length codes. Then, finer sub-sampled grids of the original image grid are progressively interpolated (i.e. their probability distributions are obtained) conditioned on previously encoded/decoded grids so far, to encode/decode all pixels of the image grid. The interpolators are neural network based and learned from a dataset.

The presented method was shown to achieve better than or on par compression performance with the recent scale-based auto-regressive models literature, yet required 10x or more less neural network parameters and encoding/decoding computation complexity. The improved compression-complexity trade-off was attributed to several contributions/findings in the overall system and neural network architecture design, such as using the same interpolator neural networks in different scales, using separate neural networks for different parameters of the probability distribution and performing the processing in the YCoCg-R color space instead of the RGB color space.

\appendices


\section*{Acknowledgment}
We would like to thank Sinem Gumus for obtaining the encoding/decoding times of several compression methods on the same computing system.

\ifCLASSOPTIONcaptionsoff
  \newpage
\fi



%
\bibliographystyle{IEEEtran}
\bibliography{thesis.bib}{} 


%







\end{document}